\documentclass[11pt,a4paper]{article}

\usepackage[a4paper]{geometry}

\usepackage[utf8]{inputenc}
\usepackage[USenglish]{babel}

\usepackage{amsmath}
\usepackage{amsfonts}
\usepackage{amssymb}
\usepackage{wasysym} 
\usepackage{siunitx}

\usepackage{enumerate}
\usepackage{graphicx}
\usepackage{xcolor}
\definecolor{red}{rgb}{1,0,0}
\definecolor{purple}{rgb}{0.7,0.2,1}

\usepackage{hyperref}
\hypersetup{
	pdftitle={Superradiance in Modified Gravity (MOG)},
	pdfauthor={Michael Florian Wondrak},
	pdffitwindow=true,
	colorlinks=true,
	linkcolor={red!50!black},
	citecolor={blue!50!black},
	urlcolor={blue!80!black}
}

\usepackage[comma,numbers,sort&compress]{natbib}
\newcommand{\inm}[1]{\ensuremath{\text{#1}}}
\newcommand{\eu}{\ensuremath{ {\text{e}} }}
\newcommand{\iu}{\ensuremath{ {\text{i}} }}
\newcommand{\piu}{\text{\ensuremath{\pi}}}
\newcommand{\dext}{\ensuremath{ {\text{d}} }}
\newcommand{\der}{\ensuremath{ d }}

\newcommand{\GN}{\ensuremath{ {G_\inm{N}} }} 
\newcommand{\Gconst}{\ensuremath{ G }} 
\newcommand{\Qg}{\ensuremath{ {Q_\inm{g}} }} 
\newcommand{\Qel}{\ensuremath{ {Q_\inm{el}} }} 
\newcommand{\Qelmax}{\ensuremath{ {Q_\inm{el,\,max}} }} 
\newcommand{\OmegaH}{\ensuremath{ {\Omega_\inm{h}} }} 
\newcommand{\OmegaHmax}{\ensuremath{ {\Omega_\inm{h}^\inm{max}} }} 
\newcommand{\rH}{\ensuremath{ {r_\inm{h}} }} 
\newcommand{\qel}{\ensuremath{ {q_\inm{el}} }} 
\newcommand{\ASc}[2]{\ensuremath{ {A^{#1}_\inm{#2}} }} 
\newcommand{\muSc}{\ensuremath{ {\mu_\inm{s}} }} 
\newcommand{\omegaSc}{\ensuremath{ {\omega} }} 
\newcommand{\kScH}{\ensuremath{ {k_\inm{h}} }} 
\newcommand{\kScI}{\ensuremath{ {k_\infty} }} 
\newcommand{\rTort}{\ensuremath{ {r_\ast} }} 
\newcommand{\Rad}{\ensuremath{R_{\omegaSc l m}}}
\newcommand{\RadT}{\ensuremath{\tilde{R}_{\omegaSc l m}}}
\newcommand{\ZSphH}{\ensuremath{Z_{\omegaSc l m}}} 
\newcommand{\SSphF}{\ensuremath{S_{\omegaSc l m}}} 
\newcommand{\KConst}{\ensuremath{K_{\omegaSc l m}}} 
\newcommand{\VPot}{\ensuremath{V_{\omegaSc l m}}} 

\newcommand{\mel}{\ensuremath{ m_\inm{e}} }		

\author{%
Michael~F.~Wondrak$^{a,b,}$%
\thanks{E-mail: \texttt{wondrak@fias.uni-frankfurt.de}} \quad %
Piero~Nicolini$^{a,b,}$%
\thanks{E-mail: \texttt{nicolini@fias.uni-frankfurt.de}} \quad %
John~W.~Moffat$^{c,}$%
\thanks{E-mail: \texttt{jmoffat@perimeterinstitute.ca}}\\[1ex]
\small $^a$ Frankfurt Institute for Advanced Studies (FIAS)\\[-0.5ex]
\small Ruth-Moufang-Str.~1, 60438 Frankfurt am Main, Germany\\[1ex]
\small $^b$ Institut f\"{u}r Theoretische Physik, Johann Wolfgang
Goethe-Universit\"{a}t Frankfurt\\[-0.5ex]
\small Max-von-Laue-Str.~1, 60438 Frankfurt am Main, Germany\\[1ex]
\small $^c$ Perimeter Institute for Theoretical Physics\\[-0.5ex]
\small 31 Caroline St N, Waterloo, Ontario N2L 2Y5, Canada
}
\date{}
\title{Superradiance in Modified Gravity (MOG)}

\begin{document}
\maketitle

\begin{abstract}
\noindent{\small%
We consider the case of rotating black holes in a dark-matter-emulating theory of gravity called MOG. The latter introduces a gravitational vector field with an associated gravitational charge proportional to the black hole mass and a scalar field in place of the gravitational constant. The resulting black hole metrics resemble the Kerr-Newman geometry and enjoy superradiant scattering. MOG, however, presents important new features. By studying the scattering of a scalar field, we show that there is a marked reduction of the critical frequency of mode amplification. This corresponds to saying that the superradiance peak frequency is red shifted. Analyses of the reflected energy flux also show that MOG black holes are fainter with respect to the standard ones. The proposed results pave the way for testing  MOG against astronomical observations.
%



}
\end{abstract}


\thispagestyle{empty}
\newpage

\section{Introduction}
General relativity (GR) is known to be an incomplete theory. It breaks down at short length scales, where quantum effects have necessarily to be taken into account. It also fails to be predictive at large length scales,  unless one invokes the presence of dark components, namely  forms of matter and energy that escape direct detection.

In such a context, black holes  might represent one of the privileged testbeds for GR
due to 
their wide mass spectrum, ranging  from Planckian values $\sim \num{e-8}$ kg to $\sim \num{e41}$ kg
in case of supermassive black holes. The black hole formation mechanism is, however, not completely understood. For instance, microscopic black holes are hypothetical objects formed by the collapse of high density fluctuations of the early Universe \cite{CaH74} or via quantum mechanical decay of de Sitter space \cite{MaR95,BoH96,MaN11}. 
The formation of black holes with intermediate mass, \textit{i.e.},
$\num{e2}$--$\num{e5}\, M_\odot$, 
is still debated: they are too heavy to be the result of a collapsing star and too light to form via an accretion mechanism. 
Due to the above uncertainties,  black holes have been the subject of investigations within theories alternative to Einstein gravity, mostly for what concerns its ultraviolet completion -- see for instance \cite{BoR00,NSS06b,Nic09,NiW11,Mof11,MMN11,IMN13,CMN15,FKN16,CNR18,Nicolini2018}.

In the present paper we want to use black holes to scrutinize the other regime, namely the large scale behavior of the gravitational field.
Specifically,  we focus on a theory called Modified Gravity (MOG). Rather than invoking little known dark components to amend GR predictions, MOG involves a massive vector field as an additional gravitational force mediator on top of the metric field, and constants of the ordinary theory are replaced by scalar fields \cite{Moffat2006}. For this reason MOG is  also known as Scalar-Tensor-Vector Gravity (STVG) theory. MOG possesses an action principle formulation, it is generally covariant, and obeys the weak equivalence principle. On the phenomenological side, MOG is able to explain observations which are typically inferred to support the dark matter paradigm~\cite{MoffatR2013,MoffatR2014,BrownsteinM2007,IsMoffat18,MoffatT2010,DeD17}. In addition, it turns out that MOG complies with cosmological observations such as the cosmic microwave background (CMB) angular power spectrum~\cite{Moffat2015c}. The theory passes the test of gravitational wave signals~\cite{Moffat2016,GreenMT2018} and predicts  gravitational lensing of galaxies and clusters~\cite{MofT09,MofRT18}.

It has been recently shown that MOG provides compelling predictions for the black hole shadow \cite{Moffat2015b,GOY18,Wang18} and the ring down of black hole mergers \cite{MMM17,ManfrediMM2018}.
Further studies about MOG black holes are presented in \cite{HussainJ2015,Moffat2016,MureikaMF2016,John2016,LopezArmengolR2017a,%
LopezArmengolR2017b,PerezLR2017,LeeH2017,SharifS2017,Pradhan2017,%
SheoranHN2018,WeiL2018,LiangWL2018}. 

In the following, we will investigate the case of black hole superradiance~\cite{Misner1972,Press1972,Zeld1971,Zeld1972,FinsterKSY2009}. Superradiance is a phenomenon consisting in the amplification of an incident wave scattering off a black hole. 
Classically, superradiance allows for energy extraction, being related to the Penrose~\cite{Penrose69} and the Blandford-Znajek~\cite{Bland77,Wilson-GerowR2016} processes. At quantum level, superradiance corresponds  a stimulated emission~\cite{Staro73,Unr74SR} -- see also \cite{Brito15}.
Apart from the phenomenology, superradiance also plays a key role within the Kerr/CFT correspondence \cite{BredbergHSS2010}.

As a general question we want to address the repercussion of MOG on the radiation amplification. In other words, we aim to clarify whether the departure from GR makes black holes fainter or brighter. 

The paper is organized as follows: 
In Sec.~\ref{sec:BH_in_MOG}, we present static and rotating black hole solutions in MOG theory. We introduce the electrically charged and rotating MOG black hole. In Sec.~\ref{sec:Scalar_Kerr-MOG}, we consider a scalar field in the Kerr-MOG spacetime and separate its equation of motion. Using appropriate boundary conditions, we derive in Sec.~\ref{sec:Superradiance} the critical frequency for superradiant scattering and we investigate its dependency on the electric charge. 
We also compare the energy flux ratios of superradiantly scattered scalar waves at the critical frequency and we explore the power ratio for a thermal incident spectrum. We summarize our findings in Sec.~\ref{sec:Summary&Outlook}.

Throughout this paper, we consider a 4-dimensional spacetime, apply the particle physics metric signature $(+,-,-,-)$, and denote the covariant derivative with respect to the metric $g_{\mu\nu}$ by $\nabla_\mu$. We use natural units, $\inm{c} \equiv \hbar \equiv 1$, but keep $\GN$ explicitly; for electrodynamic quantities we adopt Gaussian units.

\section{Black Holes in MOG}
\label{sec:BH_in_MOG}
In MOG, the Einstein-Hilbert action for the gravitational sector is amended by additional dynamical scalar and vector fields \cite{Moffat2006,MoffatT2009}: While the massive vector field $\phi^\mu$ leads to a gravitational repulsion of finite range, its mass $\mu$ and coupling parameter $\omega$ are scalar fields such that they can assume spacetime-dependent values. Additionally, the gravitational constant $G$ is a scalar field, too, which can be parametrized in terms of the MOG parameter $\alpha(x)$, $\Gconst=\GN (1+\alpha)$, where the Newtonian gravitational constant $\GN$ is effectively perceived by a test particle at distance $r \ll \mu^{-1}$ from a massive object. Thus $\alpha$ is a measure for the relative difference in the gravitational constants,
\begin{equation}
\alpha 
= \frac{\Gconst-\GN}{\GN}.
\end{equation}
 In the weak field approximation for $r \ll \mu^{-1}$, the acceleration  of a test particle reads \cite{MoffatR2013,RaM18}
\begin{equation}
\mathfrak{a}\approx  -\GN \frac{M}{r^2} \left(1 +\alpha \mu^2 r^2\right).
\end{equation}
At length scales of the Solar System $\sim 10^{12}$ m, MOG matches the Netwonian limit with an excellent accuracy since $\alpha$ is at the most $\sim 10$ and the value of  $\mu$ has to be set around $\sim (10^{20}$ m$)^{-1}$ to emulate the dark matter. Also the advancement of the Mercury perihelion has been found to be in agreement with GR \cite{Moffat2006}.

In the following, we assume a vanishing cosmological constant $\Lambda$ and discard self-interactions of the vector and scalar fields. A numerical solution to the field equations for a spherically symmetric spacetime with central Komar mass $M$ has been presented in \cite{MoffatT2009}. It shows that the scalar fields stay approximately constant, especially for large distances $r \gg 2\, \GN M$.



\subsection{Setup}
\label{subsec:setup}
In order to investigate black holes in MOG, the field equations are simplified by the assumption of constancy of the scalar fields~\cite{Moffat2015a}. The scalar field for the vector coupling is fixed to be $\omega \equiv 1$.
Observational evidence from galaxies and galaxy clusters implies a low mass of the vector field of 
$\mu \sim \SI{e-28}{\eV}$, 
which corresponds to a characteristic range of the vector interaction of 
$\mu^{-1} \sim \SI{24}{\kilo pc}$
\cite{MoffatR2013,MoffatR2014,Moffat2015a}. There are two reasons to neglect the vector field mass, $\mu \equiv 0$, in the context of black hole studies: On the one hand, it is almost vanishing in comparison with the mass of a compact object. 
On the other hand, asymptotically flat black hole geometries can realistically describe the gravitational field only locally and therefore the condition $r \ll \mu^{-1}$ is easily met.
The vector field is sourced by the so-called matter charge $\Qg$ of the black hole. As a central postulate of MOG, the black hole matter charge is proportional to Komar mass $M$~\cite{MoffatT2009},
\begin{equation}
\Qg
=\sqrt{\alpha G_N}\, M.
\end{equation}

With the assumptions above, the MOG action is given by
\begin{equation}
S_\inm{MOG}
= \frac{1}{16\piu\, G}\, \int\! \dext^4 x\, \sqrt{-g}\; R
 -\frac{1}{16\pi}\, \int\! \dext^4 x\, \sqrt{-g}\; B_{\mu\nu} B^{\mu\nu}
 +S_\inm{M}
\end{equation}
where $R$ is the Ricci scalar, $B_{\mu\nu}$ is the generalized field-strength tensor of the vector field $\phi^\mu$, 
$B_{\mu\nu} = \partial_\mu \phi_\nu -\partial_\nu \phi_\mu$, 
and $S_\inm{M}$ is the matter action. The MOG field equations in the matter-free case, \textit{i.e.}~ for vanishing matter energy-momentum tensor $T_{\inm{M}\,\mu\nu}=0$, are given by
\begin{equation}
R_{\mu\nu}
= -8\piu \Gconst\, T_{\phi\,\mu\nu}
\end{equation}
\begin{equation}
\frac{1}{\sqrt{-g}}\partial_\mu\biggl(\sqrt{-g}B^{\mu\nu}\biggr)=0
\end{equation}
\begin{equation}
\partial_\sigma B_{\mu\nu}+\partial_\mu B_{\nu\sigma}+\partial_\nu
B_{\sigma\mu}=0 .
\end{equation}
Here, $T_{\phi\,\mu\nu}$ stands for the traceless energy-momentum tensor associated with $\phi^\mu$,
\begin{equation}
T_{\phi\,\mu\nu}
=  -\frac{1}{4\piu}\, \left[
 B_\mu{}^\alpha B_{\nu\alpha}
 -\frac{1}{4}\, g_{\mu\nu}\, B^{\rho\sigma}B_{\rho\sigma}
\right]
\label{eq:timunub}
.
\end{equation}
%
%
We will also consider the case of electrically charged black holes in which $T_{\inm{M}\,\mu\nu}$ becomes the energy-momentum tensor of a Maxwell field $A_\mu$ with field-strength tensor $F_{\mu\nu}$. Its form is the same as that in \eqref{eq:timunub} with $F_{\mu\nu}$ replacing $B_{\mu\nu}$.


\subsection{Static and Rotating Black Hole Solutions}
The line element of the asymptotically flat, spherically symmetric, static MOG black hole with Komar mass $M$ reads in the Schwarzschild-like coordinates $(t,\,r,\,\theta,\,\phi)$ \cite{Moffat2006,Moffat2015a}:
\begin{align}
\label{eq:metricStatUn}
d s^2
&= \left(1 -\frac{2\, \Gconst M}{r} +\frac{\Gconst\, \Qg^2}{r^2}\right) \dext t^2
-{\left(1 -\frac{2\, \Gconst M}{r} +\frac{\Gconst\, \Qg^2}{r^2}\right)}^{-1} \dext r^2
-r^2\, \dext \Omega^2\\
&= \left(1 -\frac{2\, \Gconst M}{r} +\frac{\alpha\, \GN \Gconst\, M^2}{r^2}\right) \dext t^2
-{\left(1 -\frac{2\, \Gconst M}{r} +\frac{\alpha\, \GN \Gconst\, M^2}{r^2}\right)}^{-1} \dext r^2
-r^2\, \dext \Omega^2.
\end{align}

The rotating black hole solution in MOG depends on the spin parameter $a=J/M$ where $J$ is the Komar angular momentum of the asymptotically flat, axisymmetric, stationary spacetime. By applying Boyer-Lindquist coordinates $(t,\,r,\,\theta,\,\phi)$, the line element can be cast in a form similar to a Kerr-Newman spacetime \cite{Moffat2015a,Wald1984}:
\begin{equation}
\begin{split}
\label{eq:metricRotUn}
d s^2
= &\frac{\Delta(r)}{\rho^2 (r,\theta)}\,
{\left(\dext t -a\, \sin^2\!\theta\, \dext \phi\right)}^2
-\frac{\sin^2\!\theta}{\rho^2(r,\theta)}
{\left[\left(r^2+a^2\right)\dext \phi -a\, \dext t\right]}^2\\
&{}-\frac{\rho^2(r,\theta)}{\Delta(r)}\, \dext r^2
-\rho^2(r,\theta)\, \dext \theta^2
\end{split}
\end{equation}
with the definitions
\begin{align}
\label{eq:metricRotUn_Delta}
\Delta(r)
&\equiv r^2 -2\, \Gconst M\, r +a^2 +\alpha\, \GN\Gconst M^2,\\
\label{eq:metricRotUn_rho}
\rho^2(r,\theta)
&\equiv r^2+a^2\,\cos^2\!\theta.
\end{align}

We can obtain the electrically charged version of the black hole metric by replacing $\Qg^2$ by $\Qg^2 +\Qel^2$, 
where $\Qel$ denotes the electric charge. The only change occurs in the definition of~$\Delta$:
\begin{equation}
\label{eq:metricRotCh_Delta}
\Delta(r)
\to \Delta_\inm{el}(r)
= r^2 -2\, \Gconst M\, r +a^2 +\alpha\, \GN\Gconst M^2 +\Gconst\, \Qel^2.
\end{equation}

The outer (``h'') and inner (``$-$'') black hole horizons lie at
\begin{equation}
\label{eq:HorPosition}
r_{\inm{h}/-}
= \left(1 +\alpha\right) \GN M
\left(1
 \pm \sqrt{\frac{1}{1 +\alpha} -\frac{a^2}{{\left(\left(1 +\alpha\right) \GN M \right)}^2}
  -\frac{\Qel^2}{\left(1 +\alpha\right) \GN M^2}}
\right).
\end{equation}
In the extremal case both horizons merge at
$\rH = \left(1 +\alpha\right) \GN M$.
The allowed range for the spin parameter to ensure a hidden singularity turns to be
\begin{equation}
\label{eq:SpinParamRange}
0 \leq a \leq \sqrt{1 +\alpha}\, \sqrt{1 -\frac{\Qel^2}{\GN M^2}}\, \GN M.
\end{equation}
We note that contrary to what happens in GR the spin parameter can exceed $\GN M$.

In general, we do not expect a black hole to have substantial electric charge. One can see this by considering the ratio between electrostatic and gravitational forces. In case of protons it is of order $\sim \num{e36}$ and in case of electrons of order $\sim \num{e42}$.
A charged black hole favors the accretion of matter of opposite charge. In the case of matter of the same charge, the electric repulsion could exceed the gravitational attraction if the black hole charge was large enough. This situation would result in a charge-selective accretion eventually discharging the black hole.
Whether charge-dependent or -independent accretion occurs can be estimated by comparing the gravitational and the electrostatic potential energies of a test particle of mass $m$ and charge $\qel$. Like in GR, no discharge happens if the MOG black hole charge-to-mass ratio satisfies
\begin{equation}
\label{eq:maxChargeMassRatio}
\frac{\Qel}{\sqrt{\GN}\, M}
\leq \frac{\sqrt{\GN}\, m}{\qel}.
\end{equation}
For an estimation of the upper limit on $\Qel/M$ one has to specify the lightest free charged particles around the black hole. If we assume a hydrogen plasma consisting of protons ($m/\qel =\num{e-18}~{\GN}^{-1/2}$) and electrons ($m/\qel =\num{e-21}~{\GN}^{-1/2}$), we find $\Qel/M \leq \num{e-18}~{\GN}^{1/2}$ for a positively charged black hole and even less for a negatively charged one~\cite{Nicolini2018,Wald1984,Pag06}. 
For microscopic black holes, one finds even more stringent limits due to quantum effects. An evaporating black hole will discharge thermally and the above relation becomes in such a case
\begin{equation}
\frac{\Qel}{\sqrt{\GN}\, M}
\leq \frac{\sqrt{\GN}\, m_\inm{e}}{e}\, \left( \GN\, m_\inm{e}\, M\right),
\end{equation}
with  $M < 1/\GN m_\inm{e} \sim \SI{e14}{kg}$, where $\mel$ is the electron mass and $e$ is the elementary charge \cite{Carter1974,Gibbons1975}.

\section{Wave Scattering by MOG Black Holes}
\label{sec:Scalar_Kerr-MOG}
In this Sec. we turn towards scattering of fields by rotating black holes. For simplicity we focus on a scalar field.
Since the rotating MOG black hole spacetimes are similar to Kerr-Newman geometries, we proceed in agreement to \cite{Winstanley2001}.

\subsection{Equation of Motion}
An electrically neutral scalar field $\Phi$ of mass $\muSc$ is subject to the Klein-Gordon equation
\begin{equation}
\label{eq:KGEq}
\left(\nabla_\alpha \nabla^\alpha +\muSc^2\right)\, \Phi(t,r,\theta,\phi)
= 0.
\end{equation}
In the rotating MOG black hole spacetimes above, this equation allows for a natural separation \cite{BrillCPFI1972,GaKo}; we focus on a single monochromatic mode
\begin{equation}
\Phi(t,r,\theta,\phi)
= \Rad(r)\; \ZSphH(\theta,\phi)\; \eu^{-\iu\omegaSc t}
\end{equation}
where $\omegaSc$ is the frequency of the scattering field as measured by an observer at infinity, $\Rad(r)$ is the radial function, and $\ZSphH(\theta,\phi)$ denotes an oblate spheroidal harmonic which can be expressed in terms of the oblate spheroidal wave function $\SSphF(\theta)$,
\begin{equation}
\ZSphH(\theta,\phi)
= \SSphF(\theta)\, \eu^{\iu m \phi}.
\end{equation}
The degree $l \geq 0$ and order $-l \leq m \leq l$ specify the solid angular dependency.
The partial differential equation \eqref{eq:KGEq} splits into two ordinary differential equations with the separation constant $\KConst$:
\begin{subequations}
\begin{align}
&\frac{\der}{\der r}\, \left(\Delta(r)\, \frac{\der \Rad(r)}{\der r}\right)
+\left(
 \frac{{\left(\left(r^2 +a^2\right)\omegaSc -a\, m\right)}^2}{\Delta(r)}
 -\muSc^2 r^2 -\KConst
\right) \Rad(r)
= 0
\label{eq:ODE_rad}\\
\begin{split}
&\frac{1}{\sin\theta}\, \frac{\der}{\der \theta}\,
 \left(\sin\theta\; \frac{\der \SSphF(\theta)}{\der \theta}\right)
 +\left(-{\left(a\, \omegaSc\, \sin\theta -\frac{m}{\sin\theta} \right)}^2
 -a^2\, \muSc^2\,\cos^2\!\theta +\KConst\right) \SSphF(\theta)\\&
 = 0
\end{split}
\end{align}
\end{subequations}

In the following we concentrate on the radial equation \eqref{eq:ODE_rad}. It can be simplified by introducing a Regge-Wheeler-like coordinate $\rTort$ fulfilling
\begin{equation}
\frac{\der}{\der r}
= \frac{r^2 +a^2}{\Delta(r)}\, \frac{\der}{\der \rTort}
\end{equation}
which becomes $\rTort \to \infty$ at infinity and $\rTort \to -\infty$ at the horizon. Moreover, we define a modified radial function $\RadT(\rTort)$ by
\begin{equation}
\Rad(r)
= {\left(r^2 +a^2\right)}^{-1/2}\, \RadT(\rTort).
\end{equation}

Then the equation takes on a Schr\"odinger-like form
\begin{equation}
\label{eq:ODE_rad_Tort}
\frac{\der^2}{\der \rTort^2}\, \RadT(\rTort)
+\VPot(r)\, \RadT(\rTort)
= 0
\end{equation}
with the scattering potential
\begin{equation}
\label{eq:PotT}
\begin{split}
\VPot(r)
= &{\left(\omegaSc -\frac{a\, m}{r^2 +a^2}\right)}^2
-\frac{\Delta(r)\, \left(\muSc^2 r^2 +\KConst\right)}%
 {{\left(r^2 +a^2\right)}^2}\\
&+\frac{\Delta(r)\, \left(-2\,\Gconst M r +2\, a^2 +2\,\alpha\, \GN\Gconst M^2\right)}{{\left(r^2 +a^2\right)}^3}
-\frac{3\, a^2\, \Delta^2(r)}{{\left(r^2 +a^2\right)}^4}.
\end{split}
\end{equation}

\subsection{Asymptotic Solutions}
The boundary conditions play a crucial role in the investigation of scattering processes. They are imposed at null infinity and at the horizon. We stress here that the limit at infinity is meant to be in agreement with the asymptotic ordering of the parameters of the theory. In other words, we consider infinity to be equivalent to having $r \gg r_{\inm{h}}$ with the auxiliary condition $r \ll \mu^{-1}$ for an almost vanishing mass parameter $\mu\approx 0$.
 After such clarification, we can display the asymptotic  behaviors of the potential in eq.~\eqref{eq:PotT}. We find constant values, both at infinity 
\begin{equation}
\lim_{r\to\infty} \VPot(r)
= \omegaSc^2 -\muSc^2 \times \lim_{r\to\infty}\, \frac{r^2\, \Delta(r)}%
{{\left(r^2 +a^2\right)}^2}
= \omegaSc^2 -\muSc^2
= \left(\kScI\right)^2
\end{equation}
and at the horizon
\begin{equation}
\lim_{r\to\rH} \VPot(r)
= {\left(\omegaSc -m\,\OmegaH\right)}^2
= \left(\kScH\right)^2.
\end{equation}
Here we used the horizon angular velocity
\begin{equation}
\label{eq:Horizon_AngularVelocity}
\OmegaH
= -{\left.\frac{g_{t\phi}}{g_{\phi\phi}}\right|}_{r=\rH}
= \frac{a}{\rH^2 +a^2},
\end{equation}
and introduced the definitions
\begin{equation}
\kScH
\equiv \omegaSc -m\, \OmegaH, \quad\quad
\kScI
\equiv \sqrt{\omegaSc^2 -\muSc^2}.
\label{eq:kappas}
\end{equation}
At this point one can check the accuracy of the proposed approximation for the value of $\mu$. First, one can see that the scattering dynamics is  almost  insensitive  to  the  far-distance  regime. To appreciate the corrections coming from the breakdown of the condition $\mu\approx 0$, one can estimate the upper bound of the relative difference between the scattering potential  \eqref{eq:PotT} at $r = \mu^{-1}$ and $r\to \infty$. For  a  typical  solar-mass  black  hole  we  find such a difference is of the order $\sim 10^{-16}$, while for  a supermassive  black  hole, \textit{e.g.},   Sgr  A*,  is of the order  $\sim 10^{-9}$. Furthermore for the specific case of superradiance, the phenomenon depends on the horizon only, namely on those values of $\OmegaH$ that make $\kScH$ negative.
As a result, we can safely continue our analysis. 

By solving eq.~\eqref{eq:ODE_rad_Tort} asymptotically, one obtains the radial function:
\begin{equation}
\RadT(\rTort)
\to \begin{cases}
\ASc{\inm{h}}{in}\, \eu^{-\iu \kScH\, \rTort}
+ \ASc{\inm{h}}{out}\, \eu^{\iu \kScH\, \rTort}
& \text{for $\rTort \to -\infty$}\\
\ASc{\infty}{in}\, \eu^{-\iu \kScI\, \rTort}
+ \ASc{\infty}{out}\, \eu^{\iu \kScI\, \rTort}
& \text{for $\rTort \to +\infty$}
\end{cases}
\end{equation}
The full solution with $\rTort = \rTort(r)$ reads:
\begin{equation}
\label{eq:fullsolution}
\begin{split}
&\Phi(t,r,\theta,\phi)\\
&\to \SSphF(\theta) \times
\begin{cases}
\frac{1}{\sqrt{\rH^2 +a^2}} \left(
 \ASc{\inm{h}}{in}\,
  \eu^{-\iu (\kScH\, \rTort +\omegaSc t -m \phi)}
 + \ASc{\inm{h}}{out}\,
  \eu^{\iu (\kScH\, \rTort -\omegaSc t +m \phi)}
\right)
& \text{for $r \to \rH$}\\
\frac{1}{r} \left(
 \ASc{\infty}{in}\,
  \eu^{-\iu (\kScI\, \rTort +\omegaSc t -m \phi)}
 + \ASc{\infty}{out}\,
  \eu^{\iu (\kScI\, \rTort -\omegaSc t +m \phi)}
\right)
& \text{for $r \to \infty$}
\end{cases}
\end{split}
\end{equation}
The indices of the amplitudes $A$ determine the incident (``in'') or scattered (``out'') parts of the wave at infinity (``$\infty$'') or at the horizon (``h'').

Since our scattering field is electrically neutral, there is no coupling to a possible electromagnetic vector potential. Thus the calculation above is independent of a possible electric charge of the black hole. The only difference occurs in the value of the horizon angular velocity defined in eq.~\eqref{eq:Horizon_AngularVelocity} due the change of the horizon position.

\section{Superradiance}
\label{sec:Superradiance}

We are interested in the reflection coefficient
$\mathcal{R} \equiv \ASc{\infty}{out}/\ASc{\infty}{in}$
which indicates the amplitude ratio of the reflected and the incident radiation as measured by an observer at infinity. The effect of amplification of the radiation scattered off the black hole, \textit{i.e.}~
reflectance ${\left|\mathcal{R}\right|}^2 > 1$, is called superradiance \cite{Misner1972,Press1972,Zeld1971,Zeld1972}. We demand the boundary condition $\ASc{\inm{h}}{out}=0$, namely no outgoing radiation at the horizon as seen by any local observer, in order to ensure causality \cite{Teuko73}. Such a condition for $\ASc{\inm{h}}{out}$ determines the profile of the  so-called IN-mode in \eqref{eq:fullsolution}. 


\subsection{Superradiant Frequency Range}
In order to relate the amplitudes of the mode given in eq.~\eqref{eq:fullsolution} at the horizon and at infinity, we study the structure of the radial equation \eqref{eq:ODE_rad_Tort}. According to Abel's identity, the Wronskian of two solutions to this ordinary differential equation does not depend on the radial position. Furthermore, since the coefficients of the ordinary differential equation are real, these two solutions can be chosen to be complex conjugates. We obtain
\begin{equation}
\label{eq:RelatingAmplitudes}
\kScH\,
\left( {\left|\ASc{\inm{h}}{out}\right|}^2
 -{\left|\ASc{\inm{h}}{in}\right|}^2 \right)
= \kScI\,
\left( {\left|\ASc{\infty}{out}\right|}^2
 -{\left|\ASc{\infty}{in}\right|}^2 \right).
\end{equation}
With the help of the transmission coefficient
$\mathcal{T} \equiv \ASc{\inm{h}}{in}/\ASc{\infty}{in}$ and the boundary condition $\ASc{\inm{h}}{out}=0$, we rewrite eq.~\eqref{eq:RelatingAmplitudes} as
\begin{equation}
\label{eq:coeffrt}
{\left| \,\mathcal{R}\, \right|}^2
= 1 -\frac{\kScH}{\kScI}\, {\left| \,\mathcal{T}\, \right|}^2.
\end{equation}
Thus superradiance occurs if $\kScH < 0$ which translates into a critical frequency for the scalar field:
\begin{equation}
\label{eq:omegaScCrit}
\frac{\omegaSc}{m} < \OmegaH = \frac{a}{\rH^2 +a^2}
\end{equation}
This effect only shows up for positive orders $m \geq 1$.  This means that the rotation of the hole enhances the energy of an incoming low-frequency classical wave with an angular momentum oriented towards that of the hole. In other words, supperradiance slows down the black hole rotation \cite{BiD82}. For negative-frequency modes, superradiant scattering obeys the same condition \eqref{eq:omegaScCrit}. In this case, $m$ has to be negative. Note that in any case the absolute value of the frequency has to be larger than the field mass, $\left|\omegaSc\right| > \muSc$, in order to ensure a propagating mode.

The black hole angular velocity $\OmegaH$ acts as the normalized critical frequency. It encodes the dependency on the MOG parameter $\alpha$, the spin parameter $a$, and the electric charge $\Qel$. We explore these relations below. Thereby, we consider $0 \leq \alpha \leq 10$ as this is the range indicated by the analysis of gravitational wave signals \cite{Moffat2016}.

\subsection{Critical Frequencies}
\begin{figure*}[!htbp]
\begin{center}
	\includegraphics[width=0.8\linewidth]{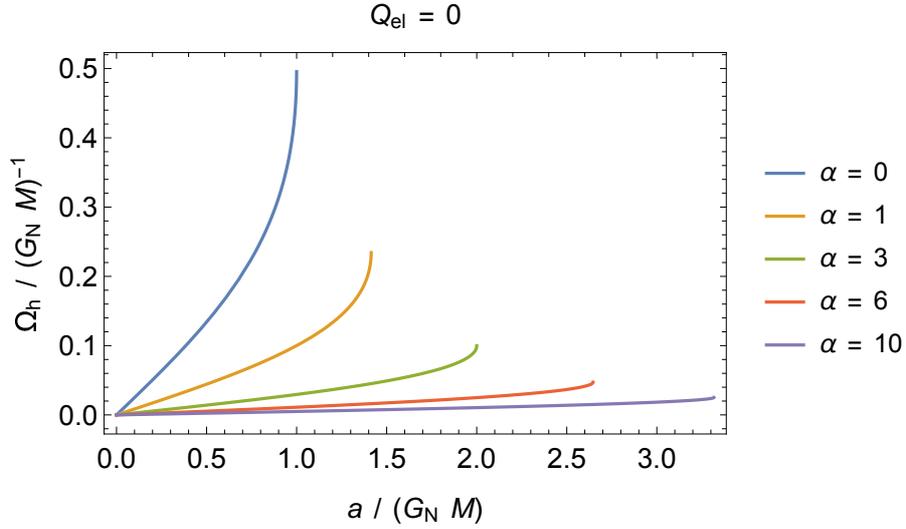}
\caption{Normalized critical frequency $\OmegaH$ as a function of the spin parameter for GR ($\alpha=0$) and MOG ($\alpha>0$). The region below a curve indicates the frequency range in which a scalar field is superradiantly scattered. Larger $\alpha$'s, $\alpha>10$, imply even smaller critical frequencies.}
	\label{fig:plotCritFrequVSa3Q0}
\end{center}
\end{figure*}

We start with electrically neutral rotating black holes in MOG. Fig.~\ref{fig:plotCritFrequVSa3Q0} shows the normalized critical frequency $\OmegaH$ as a function of the spin parameter $a$ for different values of the MOG parameter $\alpha$. The comparison of black holes of the same mass reveals that the higher is $\alpha$, the lower is  the critical frequency.
The allowed spin range for MOG black holes is extended to values $a > \GN M$ according to eq.~\eqref{eq:SpinParamRange}. As a result only in the interval $0 \leq a \leq \GN M$ a direct comparison with Kerr black holes in GR ($\alpha = 0$) is possible. 
The suppression ratio $(\OmegaH)_\mathrm{MOG}/(\OmegaH)_\mathrm{GR}$ amounts to $0.20$--$0.34$ for $\alpha = 1$ and $0.010$--$0.020$ for $\alpha = 10$. 

\begin{figure*}[!htbp]
\begin{center}
	\includegraphics[width=0.6\linewidth]{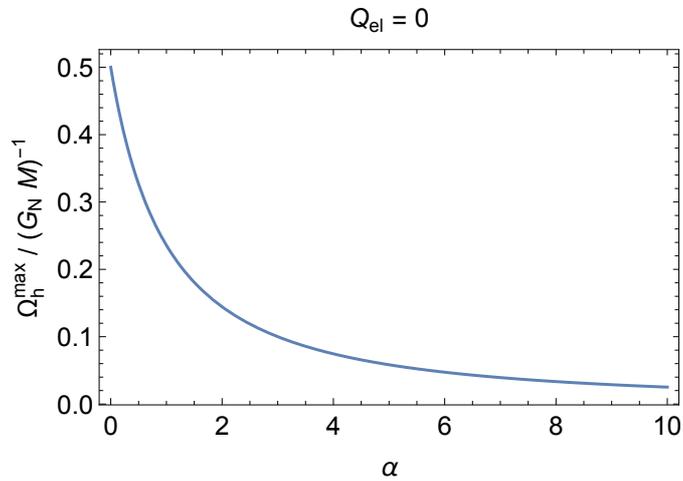}
\caption{Maximal normalized critical frequency $\OmegaHmax$ of an electrically neutral rotating black hole as a function of the MOG parameter $\alpha$. For each value of $\alpha$, the spin parameter $a$ is chosen to make the black hole extremal.}
	\label{fig:plotCritFrequExtremalVSalphaQ0}
\end{center}
\end{figure*}

 For fixed mass $M$, the maximal critical frequencies occur for extremal configurations, \textit{i.e.}~when inner and outer horizons merge. This is due to the fact that an extremal black hole possesses the smallest event horizon radius so that the horizon angular velocity becomes the largest. For what is discussed above also extremal black hole critical frequencies decrease with $\alpha$  -- see Fig.~\ref{fig:plotCritFrequExtremalVSalphaQ0}. Such a frequency decrease is more marked for $0 < \alpha \lesssim 3$. Higher values of $\alpha$ modify the frequency to a lesser extent.

\begin{figure*}[!htbp]
\begin{center}
	\includegraphics[width=0.8\linewidth]{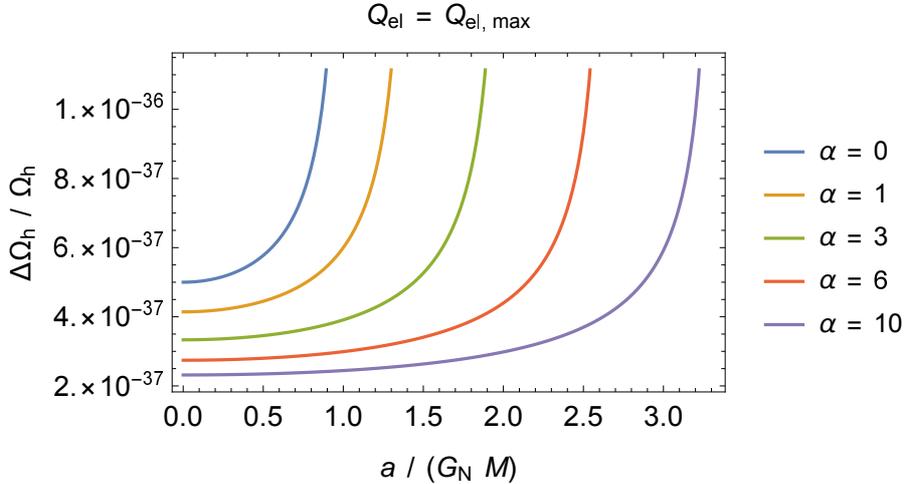}
\caption{Relative shift of the normalized critical frequency, $\Delta\OmegaH/\OmegaH$, caused by the electric charge $\Qel$ as a function of the spin parameter $a$. The largest charge allowing a stable configuration in the astrophysical context is chosen, $\Qelmax = \num{e-18}~{\GN}^{1/2} M$, from \eqref{eq:maxChargeMassRatio}.}
	\label{fig:plotCritFrequFullShiftVSaQQmax}
\end{center}
\end{figure*}

The presence of black hole electric charge modifies the horizon radius according to \eqref{eq:HorPosition}. This is the only modification affecting the scattering properties of the neutral scalar field $\Phi$ which otherwise is insensitive to the charge.
Fig.~\ref{fig:plotCritFrequFullShiftVSaQQmax} shows the relative shift of the normalized critical frequency $\OmegaH$ as a function of the spin parameter $a$. We have chosen the maximal electric charge according to eq.~\eqref{eq:maxChargeMassRatio}.

\begin{figure*}[!htbp]
\begin{center}
	\includegraphics[width=0.6\linewidth]{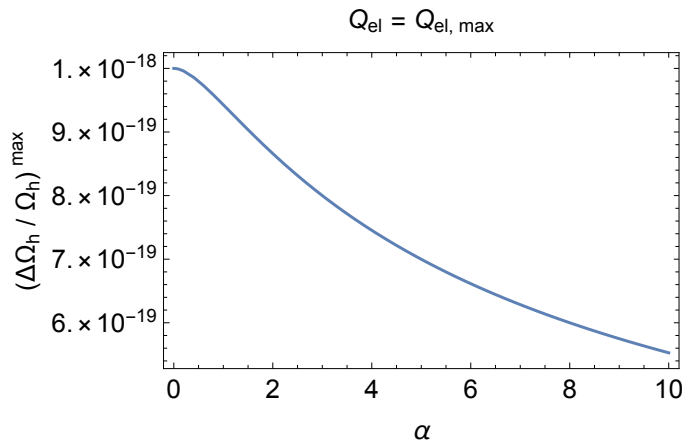}
\caption{Maximal relative shift of the normalized critical frequency due to electric charge of the black hole as a function of the MOG parameter $\alpha$. For this case, only extremal black holes with the largest astrophysically reasonable charge, $\Qelmax = \num{e-18}~{\GN}^{1/2} M$, are considered -- see  \eqref{eq:maxChargeMassRatio}.}
	\label{fig:plotCritFrequFullShiftMaxVSalphaQQmax}
\end{center}
\end{figure*}

We note that the electric charge has an antagonist effect to MOG: It enhances the critical frequency. The largest contribution arising for extremal black holes (which have a smaller spin cf.~eq.~\eqref{eq:HorPosition}) is displayed in Fig.~\ref{fig:plotCritFrequFullShiftMaxVSalphaQQmax}. However, since the maximal astrophysical charge-to-mass ratio is so small, the effects on a neutral scalar field are negligible. In the following, we restrict our attention to neutral black holes.


\subsection{Energy Flux Ratios}
As a direct consequence of a smaller value of $\OmegaH$, MOG manifests itself in a suppression of higher frequencies of the reflected wave spectrum. This means that only a smaller range of the initial frequency spectrum can be amplified. This attenuation has an even larger impact on the luminosity since the superradiant frequencies are less energetic than the absorbed ones.
The outgoing energy flux $\dot{E}$ measured at infinity, namely for $\rH\ll r\ll \mu^{-1}$ as outlined above, can be calculated from the field's energy-momentum tensor and by considering an infinitesimal timelike surface element of a constant $r$-slice at infinity. In the case of a monochromatic massive scalar field, one can derive (cf.~also \cite{Brito15})
\begin{equation}
\dot{E}
= \frac{\omegaSc\, \kScI}{2}\, {\left|\ASc{\infty}{out}\right|}^2
= \frac{\omegaSc\, \kScI}{2}\, {\left| \,\mathcal{R}\, \right|}^2\, 
 {\left|\ASc{\infty}{in}\right|}^2,
\end{equation}
where $\mathcal{R}$ and $\ASc{\infty}{in}$ depend on the frequency $\omegaSc$, the degree $l$, and the order $m$.

\begin{figure*}[!htbp]
\begin{center}
	\includegraphics[width=0.8\linewidth]{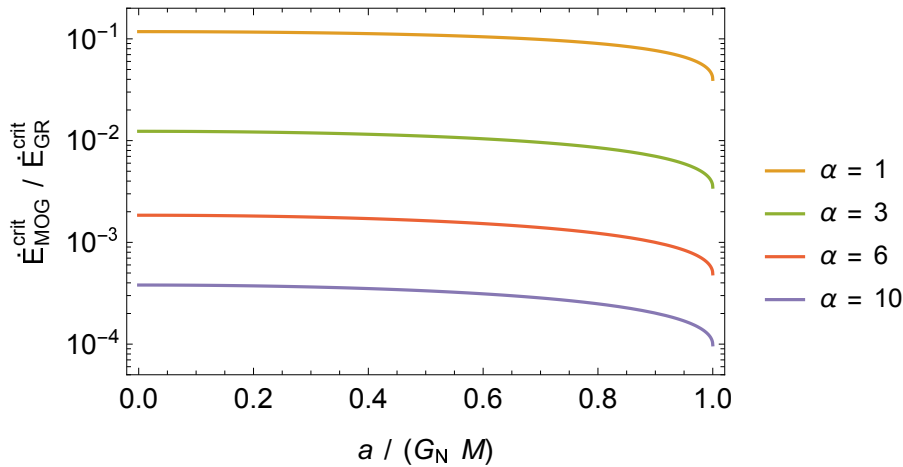}
\caption{Ratio of the outgoing energy flux at infinity as a function of the spin parameter $a$ for neutral MOG and GR black holes. In this case, the incoming waves are monochromatic with frequencies equaling the respective critical frequency, \textit{i.e.}~$(\OmegaH)_\inm{MOG}$ and $(\OmegaH)_\inm{GR}$, and $\muSc=0$.}
	\label{fig:plotFluxRatioAtCritVSaQ0Log}
\end{center}
\end{figure*}

As a start, we compare the MOG and GR energy fluxes of two monochromatic waves with the same ingoing amplitude $\ASc{\infty}{in}$, but at the respective critical superradiant frequency. Since, by definition, $\left| \,\mathcal{R}\, \right|_{\omegaSc= m\OmegaH} = 1$, the ratio becomes
\begin{equation}
\frac{\dot{E}^\inm{crit}_\inm{MOG}}{\dot{E}^\inm{crit}_\inm{GR}}
= \frac{(\OmegaH)_\inm{MOG}\, (\kScI)_\inm{MOG}}{(\OmegaH)_\inm{GR}\, (\kScI)_\inm{GR}}
\leq \frac{(\OmegaH)_\inm{MOG}^2}{(\OmegaH)_\inm{GR}^2}
< 1,
\end{equation}
where we have used~eqs.~\eqref{eq:omegaScCrit} and \eqref{eq:HorPosition} in the last step. This ratio in the massless case $\muSc=0$ is displayed in Fig.~\ref{fig:plotFluxRatioAtCritVSaQ0Log} in the spin range applicable to GR black holes. As expected, there is a strong dependency on $\alpha$ due to the quadratic dependency on the critical frequency ratio. Already for $\alpha > 3$, the flux ratio becomes much smaller than unity.

The analysis of polychromatic wave scattering is very related to the incident spectrum and thus to the external source of radiation at infinity.
Those frequencies of the incident spectrum that are below the critical frequency of GR but above the critical frequency of MOG are superradiantly scattered by the GR black hole while absorbed by the MOG black hole. This leads to the conclusion that the MOG black hole is fainter.
The larger is the value of $\alpha$, the bigger is the difference in the respective critical frequencies, and thus the larger becomes the frequency range of the incident spectrum for which MOG black holes will not superradiate. 

To better clarify this result, we offer an example with a specific incident spectrum. To compare the MOG and GR fluxes, one has to integrate over the flux contribution of each mode weighted by  the normalized initial mode distribution $n(\omega)$,
\begin{equation}
\label{eq:TotFlux}
\dot{E}_\inm{tot}
= \int\! \dext \omega\; \frac{\omegaSc\, \kScI}{2}\,
{\left|\mathcal{R}(\omega)\right|}^2\, n(\omega).
\end{equation}
While in principle waves of different degrees and orders contribute, we consider waves with $l=m=1$ which show the largest superradiant amplification \cite{Press1972,Brito15}.
For the incident radiation, we consider a thermal spectrum at the temperature $T$  as a specific case. This choice is justified in the presence of background sources such stars or the CMB radiation, despite our analysis is based on the scalar field only. We stress, however, that we do not consider a single source at a certain direction, but a bunch of emitters with an angular distribution according to the mode under consideration, \textit{i.e.}~$l=m=1$. We also consider negligible the contribution due to the Hawking temperature $T_\mathrm{H}$ of the black hole, namely 
\begin{equation}
T\gg T_\mathrm{H}
=\frac{\left( \rH -r_{-}\right)}{4\pi\, a \, k_\inm{B}}\, \OmegaH .
\end{equation}
Upon the above conditions we assume for the incident spectrum the normalized black body mode spectrum of a massless scalar field
\begin{equation}
n(\omegaSc)
= \frac{\omegaSc^2}%
{2\, \zeta(3)\, k_\inm{B}^3\, T^3\, %
\left(\exp\!\left(\frac{\omegaSc}{k_\inm{B}\, T}\right) -1\right)},
\end{equation}
where $\zeta(x)$ denotes the Riemann zeta function and $2\zeta(3)\simeq 2.40$. 
The peak frequency $\omega_\inm{peak}$ is related to the temperature by
\begin{equation}
\omega_\inm{peak}
= \left[2 +W\!\left(-\frac{2}{\inm{e}^2}\right)\right] k_\inm{B} T 
\simeq 1.59\, k_\inm{B} T \, ,
\end{equation}
where $W(x)$ is the Lambert W function. Therefore, the temperature regime $T \gg T_\inm{H}$ implies  
\begin{equation}
\frac{\omega_\inm{peak}}{\OmegaH}
\gg \frac{1.59}{4\pi}\, \left(\frac{\rH -r_{-}}{a}\right)
\label{eq:omegavspeak}
\end{equation}
 which allows for $\omega_\inm{peak} < \OmegaH$ only for $\rH\sim r_{-}$, \textit{i.e.}, when the MOG black hole is close to its extremal configuration. 

\begin{figure*}[!htbp]
\begin{center}
	\includegraphics[width=0.8\linewidth]{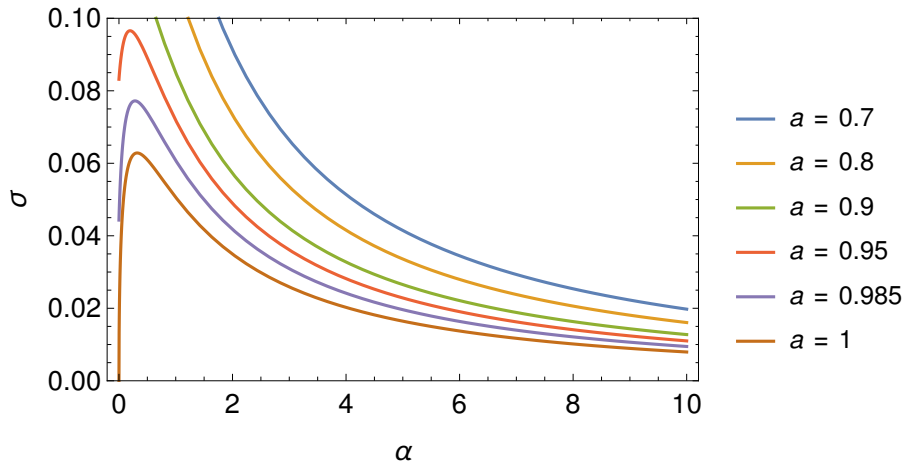}
\caption{Exclusion curve $\sigma$ for the normalized peak frequency, $\omega_\inm{peak} / (\OmegaH)_\inm{GR}$, as a function of the MOG parameter $\alpha$ for different spin parameters $a$ in $\GN M$ units. The black hole electric charge is set to zero. Larger peak frequencies $\omega_\inm{peak}$ ensure that the temperature of the incident radiation is much higher than the black hole Hawking temperature, $T \gg T_\inm{H}$.}
	\label{fig:plotSigmaVSalpha}
\end{center}
\end{figure*}

\begin{figure*}[!h]
\begin{center}
	\includegraphics[width=0.6\linewidth]{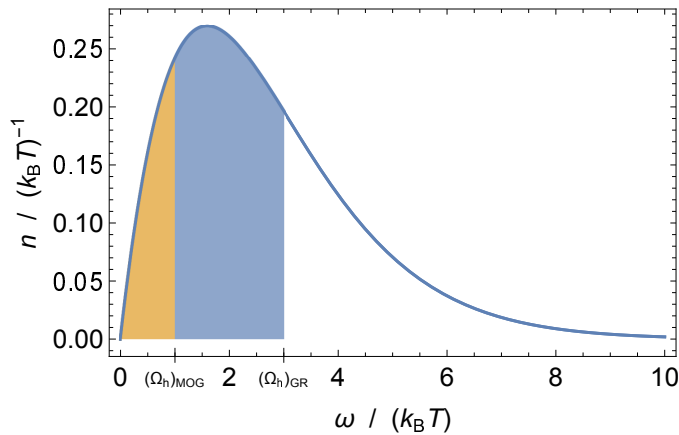}
\caption{
Normalized thermal mode distribution $n(\omega)$ as a function of the frequency~$\omegaSc$.
The orange region represents a frequency range which is superradiantly scattered in the MOG case with horizon angular velocity $(\OmegaH)_\inm{MOG}$. In contrast, the combination of the orange and blue regions stands for the frequency range superradiantly scattered by a GR black hole of the same mass and spin parameter, but a different horizon angular velocity of $(\OmegaH)_\inm{GR}$.}
	\label{fig:plotBBNormalizedModeDensityDistinctionVSomega}
\end{center}
\end{figure*}

Conversely the above condition can be easily met in the regime $\OmegaH<\omega_\inm{peak}<(\OmegaH)_\inm{GR}$.
To see this, we write \eqref{eq:omegavspeak}  as
\begin{equation}
\frac{\omega_\inm{peak}}{(\OmegaH)_\inm{GR}}
\gg \frac{1.59}{4\pi}\, \left(\frac{\rH -r_{-}}{a}\right)\left(\, 
 \frac{\OmegaH}{(\OmegaH)_\inm{GR}}\right)
\equiv \sigma
\end{equation}
where we define the function $\sigma$ that is plotted in Fig. ~\ref{fig:plotSigmaVSalpha}. The exclusion curve depends both  on the MOG and spin parameter. In the  limit $a \simeq \GN M$ and/or for $\alpha>4$, one finds $\sigma\ll 1$ corresponding to a negligible Hawking temperature even for low $\omega_\inm{peak}$, namely  in the interval $\OmegaH<\omega_\inm{peak}<(\OmegaH)_\inm{GR}$.
In Fig.~\ref{fig:plotBBNormalizedModeDensityDistinctionVSomega}, such a frequency interval is analyzed.  One can see the thermal spectrum and a sketch of the portion of frequency modes that are superradiantly scattered by a MOG and a GR black hole, respectively.
 
For the integration in eq.~\eqref{eq:TotFlux}, we need the reflectance ${\left| \,\mathcal{R}\, \right|}^2$ as a function of the frequency which has in general no analytic expression. We can however approximate the reflectance as 
\begin{equation}
{\left| \,\mathcal{R}\, \right|}^2
= \Theta\!\left(\OmegaH -\frac{\omegaSc}{m}\right),
\end{equation}
since it exhibits a plateau and a  maximum around $\sim 1.004$ in case of a scalar field in GR \cite{Press1972,Brito15}.

\begin{figure*}[!htbp]
\begin{center}
	\includegraphics[width=0.6\linewidth]{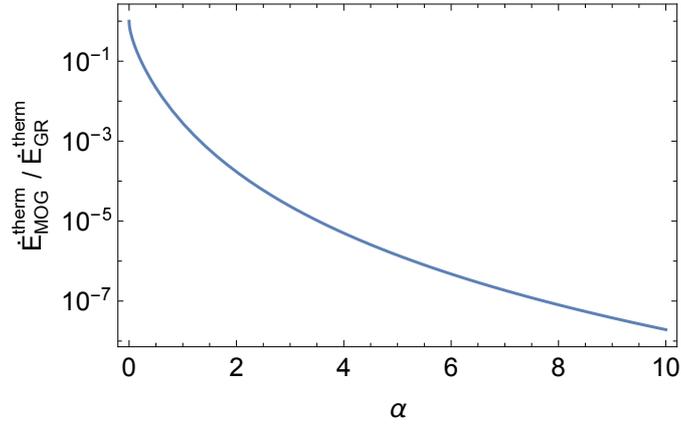}
\caption{
Superradiated energy flux ratio for an incoming thermal spectrum as a function of $\alpha$. The spin of the neutral black hole is $a = \GN M$ implying that the GR black hole is extremal with vanishing Hawking temperature $T_\inm{H}$. The peak frequency of the thermal spectrum is chosen to be $\omega_\inm{peak} = (\OmegaH)_\inm{GR}$. 
 }
	\label{fig:plotFluxRatioVSalphaLog}
\end{center}
\end{figure*}

\begin{figure*}[!htbp]
\begin{center}
	\includegraphics[width=0.8\linewidth]{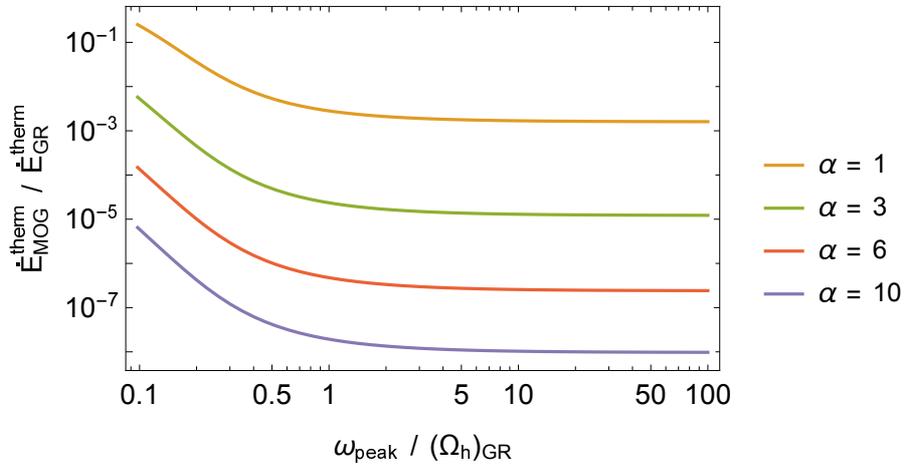}
\caption{Energy flux ratio for a thermal initial spectrum as a function of the peak frequency, $\omega_\inm{peak}$. Note that $\omega_\inm{peak}$ is proportional to the temperature of the spectrum. The spin of the neutral black hole is $a = \GN M$ implying that the GR black hole is extremal and has vanishing Hawking temperature $T_\inm{H}$.}
	\label{fig:plotFluxRatioVST3}
\end{center}
\end{figure*}

The resulting energy flux ratio as a function of $\alpha$ is displayed in Fig. \ref{fig:plotFluxRatioVSalphaLog} for a spectrum peak frequency $\omega_\inm{peak} = (\OmegaH)_\inm{GR}/2$. 
The strong suppression emerges also for small values of $\alpha$.
In Fig. \ref{fig:plotFluxRatioVST3}, the flux ratio is displayed as a function of $\omega_\inm{peak}$ for different values of $\alpha$. One can see that the MOG superradiance dies quickly off as $\omega_\inm{peak}$ approaches $(\OmegaH)_\inm{GR}$ from below. 
From the  above two cases, we conclude that the superradiance is ``switched off'' quite drastically if the MOG parameter $\alpha$ is ``turned on''.

Before the conclusion, we offer some additional considerations above the case of a massive field and the spectrum characteristics. From \eqref{eq:kappas}  one learns that not only $\omega$ is bounded from above but 
$\muSc$ too, namely 
\begin{equation}
\muSc 
< \left| \omega \right| 
< \left| m \right|\, \OmegaH
= \left| m \right|\, \frac{a}{\rH^2+a^2}
\equiv \muSc{}_\inm{,\,max}.
\end{equation}
This implies that the bigger is the black hole,  the lighter is the particle that can be superradiantly scattered.
A solar mass black hole would have $\muSc{}_\inm{,\,max} \approx \SI{6.7e-11}{eV}$ for $m=1$ in the most promising case of an extremal configuration. This implies that only massless particles can be superradiated, since such a limit for $\muSc{}_\inm{,\,max}$ is ten order of magnitude smaller than the limit on the neutrino mass. Interestingly only very soft massless particles can be superradiated. For instance the CMB is too hot for a solar mass black hole. Photons at $T=2.7$ K undergo a superradiance by a GR black hole only if its mass is smaller than $\sim \num{e24}$ kg. In case of MOG, CMB superradiance can only occur for even smaller masses.

\section{Summary and Outlook}
\label{sec:Summary&Outlook}

In this paper we have considered the problem of the superradiance for rotating black holes in the modified theory of gravity called MOG. Such a theory has as a main feature the presence of a massive vector field that mediates the gravitational interaction together with the metric. This implies that black holes have a gravitational charge proportional to their mass. Additionally the gravitational constant is a scalar field depending on a parameter called $\alpha(x)$.

By assuming the Boyer-Lindquist coordinates, the axisymmetric black hole turns to be a short-scale solution of the field equations resembling the Kerr-Newman geometry as far as the vector field mass can be neglected, $\mu\approx 0$. On the top of such a spacetime, we have studied the scattering properties of a scalar field. The phenomenon of superradiance occurs due to an enhancement of the amplitude of scattered modes, that impinge on the spinning black hole with collinear angular momentum. 

The main results emerging from MOG is a drastic reduction of the critical frequency of superradiance also for small values of the parameter $\alpha$. This implies that the reflected spectrum is red-shifted in the sense that a larger portion of incoming frequency modes are absorbed. Such a behavior has repercussions on the power spectrum. MOG black holes turn to be fainter with respect to their GR counterparts. 
From the study of a thermal bath of incoming particles we showed that only very soft massless particles can be emitted. For instance the CMB photons are too hot to be superradiantly scattered by a solar mass black hole. 
%
The case of charged superradiance confirms this scenario since the electric charge has negligible effect on the spectrum.

Despite the current analysis concerns the scalar field only, we argue that the case of higher spin cannot modify our conclusions. This is due to the fact that MOG introduces modifications of the critical frequencies which are a property of the black hole and do not depend on the nature of the scattered field. In other words, the black hole will slow its rotation and feed the field with energy to a lesser extent than in the GR case.

We believe the current work is subject to further analyses in connection to the jet launching mechanisms. We also expect that the proposed results can be tested against astronomical observations in a near future.



\section*{Acknowledgements}
The authors want to thank M.~Bleicher for valuable discussions. MFW is grateful to the Perimeter Institute for Theoretical Physics for its generous hospitality during the initial stages of this project, and acknowledges the support by the Stiftung Polytechnische Gesellschaft Frankfurt am Main. The work of PN has been supported by the project ``Evaporation of the microscopic black holes'' of the German Research Foundation (DFG) under the grant NI 1282/2-2 and by the Helmholtz International Center for FAIR within the framework of the LOEWE program (Landesoffensive zur Entwicklung Wissenschaftlich-Ökonomischer Exzellenz) launched by the State of Hesse. This research was supported in part by the Perimeter Institute for Theoretical Physics. Research at the Perimeter Institute is supported by the Government of Canada through the Department of Innovation, Science and Economic Development Canada and by the Province of Ontario through the Ministry of Research, Innovation and Science.

\providecommand{\href}[2]{#2}\begingroup\raggedright\endgroup

\end{document}